\documentclass[conference,comsoc]{IEEEtran}
\usepackage{tikz}
\usepackage{float}
\usepackage{amsmath}
\usepackage{amssymb}
\usepackage{commath}
\usepackage{algorithm}
\usepackage{algorithmic}
\usepackage{multicol}
\usepackage[noadjust]{cite}
\usepackage{balance}
\usepackage{ragged2e}
\usetikzlibrary{arrows,arrows.meta,
decorations.markings,snakes,patterns,positioning,fit,shapes}
\newcommand{\dij}[2]{\sqrt{(x_i^{#2} - x_{#1})^2 + (y_i^{#2} - y_{#1})^2}}

\newcommand{\diag}{\text{diag}}
\newcommand{\tijt}[1]{\tilde{T}_{ij}^{#1}}
\newcommand{\rijt}[1]{\tilde{R}_{ij}^{#1}}
\newcommand{\tij}[1]{T_{ij}^{#1}}
\newcommand{\rij}[1]{R_{ij}^{#1}}
\newcommand{\muest}{\boldsymbol{\mu}_{\text{est}}}
\newcommand{\mucorr}{\boldsymbol{\mu}_{\text{corr}}}
\newcommand{\mupred}{\boldsymbol{\mu}_{\text{pred}}}
\newcommand{\coest}{\mathbf{\Sigma}_{\text{est}}}
\newcommand{\cocorr}{\mathbf{\Sigma}_{\text{corr}}}
\newcommand{\copred}{\mathbf{\Sigma}_{\text{pred}}}
\newcommand{\ttat}[1]{\boldsymbol{\xi}_i^{#1}}
\newcommand{\tgam}{\frac{1}{\tilde{\gamma}_i}}
\newcommand{\tg}{\tilde{\gamma}_i}
\newcommand{\ttet}{\tilde{\theta}_i}

\newcommand{\tetvec}[1]{\boldsymbol{\xi}_{#1}}
\newcommand{\tetvect}[1]{\boldsymbol{\xi}_{#1}}

\newcommand{\zji}{\mathbf{z}_{ij}}

\newcommand{\ccij}{\mathbf{c}_{ij}}

\newcommand{\cntn}[2]{p(\ccij^{#1}, \varphi_{ij}^{#1}|\tetvec{i}^{#2})}
\newcommand{\cntnt}[2]{p(\ccij^{#1}, \varphi^{#1}|\tetvect{i}^{#2})}

\newcommand{\tntt}[2]{p(\tetvect{i}^{#1}|\tetvect{i}^{#2})}

\begin{document}
\IEEEoverridecommandlockouts
\title{Bayesian Joint Synchronization and Localization Based on Asymmetric Time-stamp Exchange
\thanks{The research leading to these results has received funding from the European Union's Framework Programme Horizon 2020 for research, technological development and demonstration under grant agreement No. 871428 (5G-CLARITY).}}

\author{\IEEEauthorblockN{Meysam Goodarzi\rlap{\textsuperscript{\IEEEauthorrefmark{1}}},\,\textsuperscript{\IEEEauthorrefmark{2}} Nebojsa Maletic\rlap{\textsuperscript{\IEEEauthorrefmark{1}}},\, Jes{\'u}s Guti{\'e}rrez\rlap{\textsuperscript{\IEEEauthorrefmark{1}}}, and Eckhard Grass\rlap{\textsuperscript{\IEEEauthorrefmark{1}}},\,\textsuperscript{\IEEEauthorrefmark{2}}}
\IEEEauthorblockA{\IEEEauthorrefmark{1}IHP -- Leibniz-Institut f\"{u}r innovative Mikroelektronik, Frankfurt (Oder), Germany}
\IEEEauthorblockA{\IEEEauthorrefmark{2}Humboldt University of Berlin, Berlin, Germany.}
Emails:\{goodarzi, maletic, teran, grass\}$@$ ihp-microelectronics.com}
\maketitle
\begin{abstract}
In this work, we study the joint synchronization and localization (sync\&loc) of Mobile Nodes (MNs) in ultra dense networks. 
 In particular, we deploy an asymmetric time-stamp exchange mechanism between MNs and Access Nodes (ANs), that,  traditionally, provides us with information about the MNs' clock offset and skew. However, information about the distance between an AN and a MN is also intrinsic to the propagation delay experienced by exchanged time-stamps. In addition, we utilize Angle of Arrival (AoA) estimation to determine the incoming direction of time-stamp exchange packets, which gives further information about the MNs' location. Finally, we employ Bayesian Recursive Filtering (BRF) to combine the aforementioned pieces of information and jointly estimate the position and clock parameters of MNs. The simulation results indicate that the Root Mean Square Errors (RMSEs) of position and clock offset estimation are kept below 1 meter and 1 ns, 
respectively.
%

\end{abstract}
%
\IEEEpeerreviewmaketitle
\section{Introduction}\label{sec:intro}
The fifth generation (5G) of mobile communication networks is expected to provide an enormous variety of localization-based services \cite{maletic2019experimental,koivisto2017joint,zheng2009joint}. User tracking \cite{wu2010clock}, next crossing cell prediction \cite{goodarzi2019next}, and location-assisted beamforming \cite{maletic2018device} can be considered as examples where Mobile Node (MN) localization plays a decisive role. State-of-the-art MN localization techniques rely primarily on the cooperation among Access Nodes (ANs), requiring them to be precisely synchronized. In addition, for many of the existing techniques to function, the clock parameters of the MNs need to be known (or to be continuously tracked). Therefore, it appears that the three aforementioned problems, namely inter-AN synchronization, MN localization, and MN's clock parameter estimation are closely intertwined and need to be addressed jointly.

In \cite{goodarzi2020synchronization,goodarzi2020hybrid}, we have thoroughly addressed the end-to-end synchronization in 5G networks. In particular, we employed Belief Propagation (BP) and Bayesian Recursive Filtering (BRF) not only to achieve high-precision end-to-end synchronization, but also to keep the inter-AN relative clock offset and skew low. In other words, the algorithms therein pave the way for the joint synchronization and localization (sync\&loc) of MNs by accurately synchronizing the neighboring ANs. 

The joint MN sync\&loc problem has been extensively considered in literature. In \cite{yuan2016cooperative}, the authors rely on symmetric time-stamp exchange between ANs and MNs to jointly and distributedly estimate MNs' location and clock offset with the aid of BP. Furthermore, the authors of \cite{etzlinger2017cooperative,meyer2018scalable} adopt a similar approach using an asymmetric time-stamp exchange mechanism proposed in \cite{chepuri2012joint}, enabling them to track both the clock offset and skew. While support of time-stamp exchange in 5G networks is a valid assumption to make (as it has been already introduced in several standards, e.g. IEEE 802.11 under the name \textit{fine time measurement} \cite{7786995}), the high number of message-passings required by BP renders the approach limited in practice. Additionally, they provide the estimation of the sync\&loc parameters at MN, whereas for the location-based services to be delivered, these parameters need to be computed on the network side.

In \cite{werner2015joint}, the authors leverage Extended Kalman Filtering (EKF) to obtain the estimation of clock parameters and position in ultra dense networks. In particular, they assume synchronized ANs and perform MN joint sync\&loc in the presence of uncertainty about Time of Arrival (ToA) and Angle of Arrival (AoA) parameters. The level of uncertainty is then determined based on the derived Cramer Rao Bound (CRB). However, in practice, the estimation accuracy of AoA and ToA plays a significant role in the performance of joint sync\&loc. Thus, a more detailed and in-depth analysis is required to recognize the limitations they impose on joint sync\&loc algorithms. Specifically, in this work, we focus on the limitations caused by uncertainty in time-stamping (which directly translates to uncertainty in ToA) while drawing on the CRB for AoA estimation and leave its detailed analysis for future works.


The contribution of this paper is summarized as follows:
\begin{itemize}
\item We present a realistic system model for joint sync\&loc based on asymmetric time-stamp exchange.
\item We propose a BRF-based joint sync\&loc algorithm using time-stamp exchange between ANs and MNs.
\item We analyze the performance of the proposed approach with the aid of detailed simulations in a challenging real world scenario. 
\end{itemize} 
The rest of this paper is structured as follows: In Section II, we introduce our system model. Section III describes the details of the BRF algorithm for joint estimation of location and clock parameters. Furthermore, simulation results are presented and discussed in Section IV. Finally, Section V concludes this work and indicates potential future work.
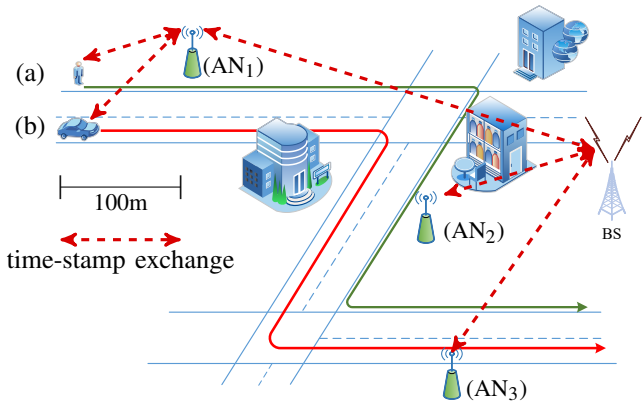
\begin{figure}[t!]
\begin{tikzpicture}[scale=1]
   \draw (0, 0) node[inner sep=0] {\includegraphics[width=0.9\linewidth]{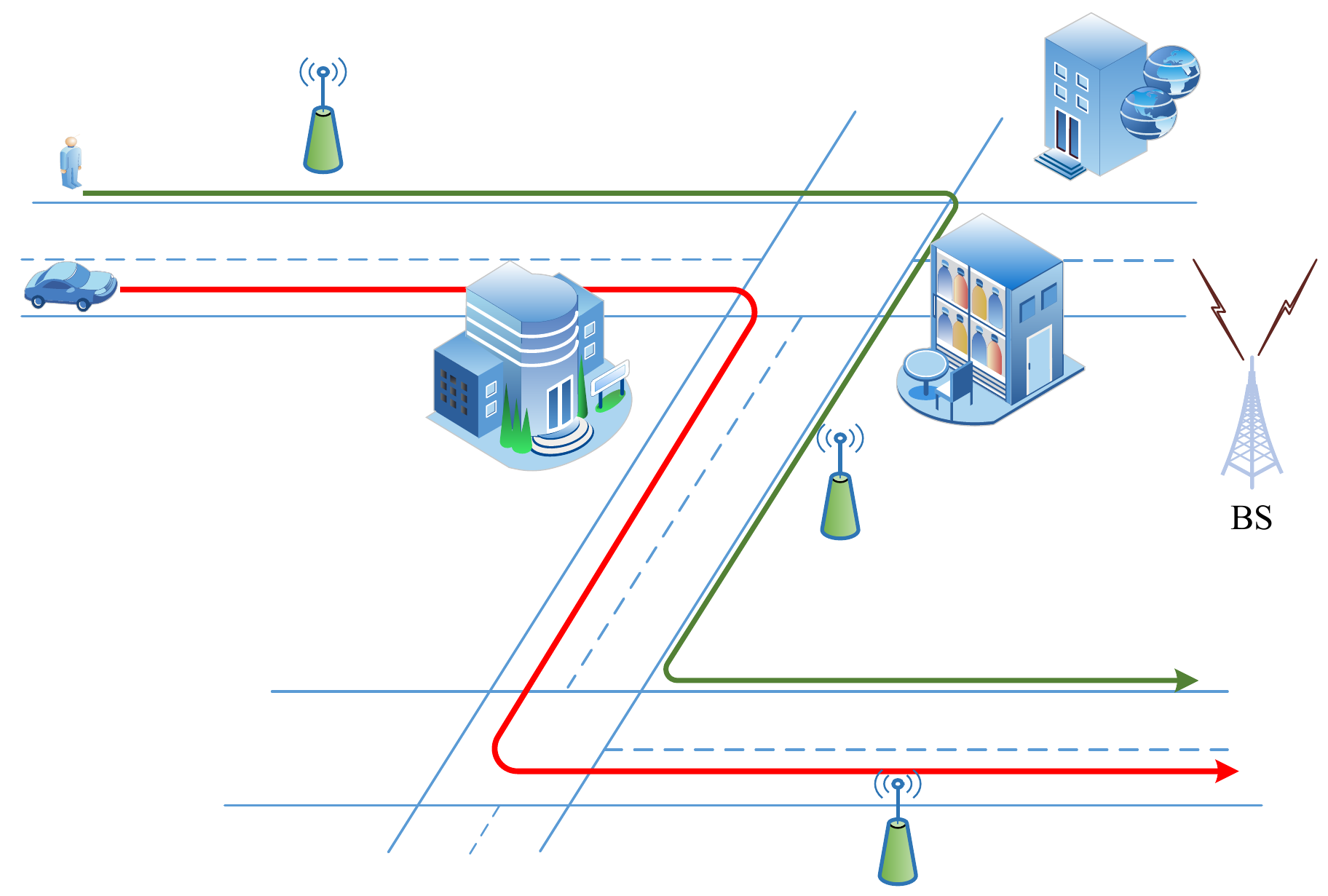}};
   \draw (-4.2,1.7) node(a0){(a)};
   \draw (-4.2,1) node(a1){(b)};
   \draw (-1.5,1.8) node(an1){\small(AN$_1$)};
   \draw (1.7, -0.4) node(an2){\small(AN$_2$)};
   \draw (2,-2.5) node(an3){\small(AN$_3$)};
   
   \draw (-3.8,0.2) node(a0){$|$};
   \draw (-2.2,0.2) node(a1){$|$};
   \draw (-3.,0.) node(a2){\small $100$m};
   
   \draw[<->,red!85!black, very thick, dashed, >=stealth'] (-1.9,2.3) node (an1){} -- (3.3, 0.7) node (bs){};
   \draw[<->, red!85!black, very thick, dashed, >=stealth'] (1.3,0.1) node (an1){} -- (3.3, 0.7) node (bs){};
   \draw[<->, red!85!black, very thick, dashed, >=stealth'] (1.4,-2.0) node (an1){} -- (3.3, 0.7) node (bs){};
   
   \draw[<->,red!85!black, very thick, dashed, >=stealth'] (-2.2,2.3) node (an1){} -- (-3.5, 1.9) node (bs){};
   \draw[<->,red!85!black, very thick, dashed, >=stealth'] (-2.2,2.3) node (an1){} -- (-3.4, 1.1) node (bs){};
   
   \draw[-] (a1.center) -- (a0.center);
   \draw[<->, red!85!black, very thick, dashed, >=stealth'] (-3.8,-.5) node (an1){} -- (-2.2, -.5) node (bs){} node[midway, below]{{\color{black}{time-stamp exchange}}};
\end{tikzpicture}
\centering
\caption{An example where MN joint sync\&loc can be carried out.}
\label{fig:scenario}
\end{figure}
\begin{table}[t!]
\centering
\caption{Notation}
\begin{tabular}{p{2cm} p{5.8cm}}
\hline
Denotation & Description \\
\hline
$\boldsymbol{A}$ & matrices \\ 
$\boldsymbol{a}$ & vectors \\ 
$\boldsymbol{a}(n)$ & $n$-th element of vector $\boldsymbol{a}$ \\ 
$\boldsymbol{I}_N$ & $N\times N$ dimensional identity matrix \\
$\boldsymbol{0}_N$ & $N\times N$ dimensional all-zero matrix \\
$\mathcal{N}(\mathbf{x}|\boldsymbol{\mu}, \boldsymbol{\Sigma})$ & Gaussian distributed random vector $\mathbf{x}$ with mean vector $\boldsymbol{\mu}$ and covariance matrix $\boldsymbol{\Sigma}$ \\
diag$(x_1, \cdots, x_K)$ &  diagonal matrix with the diagonal elements given by $(x1, \cdots, x_K)$ \\
$\thicksim$ & stands for ``is distributed as" \\ 
$\propto$ & linear scalar relationship between two real valued functions \\ \hline
\end{tabular}
\label{tab:notation}
\end{table}
\section{System Model}
We consider a scenario where a MN, e.g. a moving car/person, is served by a set of ANs, all backhauled by a Base Station (BS), as shown in Figure \ref{fig:scenario}. We assume that the ANs continuously synchronize themselves with the backhauling BS using the methods described in \cite{goodarzi2020synchronization, goodarzi2020hybrid}. The joint sync\&loc is then performed for the scenario where the MN exchanges time-stamps through an active Line-of-Sight (LoS) connection with only one AN. However, if there are further ANs in LoS to the MN, they can passively cooperate with the main AN to further enhance the performance. Moreover, an estimation of AoA is carried out upon each round of time-stamp exchange. In the following subsections, we firstly present the clock model for ANs and MNs. Then, we explain the time-stamp exchange mechanism as well as the concept of active/passive connection between ANs and MNs. Lastly, we deal with the CRB of AoA estimation.



\subsection{Clock Model}
We consider the following clock model for each node $i$.
\begin{equation}
c_i(t) = \gamma_i t + \theta_i,
\label{eq:clkmod}
\end{equation}
where $t$ represents the reference time. Furthermore, $\gamma_i$ and $\theta_i$ denote the clock skew and offset, respectively. The parameter $\gamma_i$ is generally random and varies over time. However, it is common to assume that it remains constant in the course of one synchronization period \cite{etzlinger2014cooperative,leng2011distributed,giorgi2011performance}. Given that, the goal of time synchronization can be defined as the estimation of $\gamma_i$ and $\theta_i$ (or transformations thereof) for each node.
\begin{figure}[t!]
\includegraphics[width=0.9\linewidth]{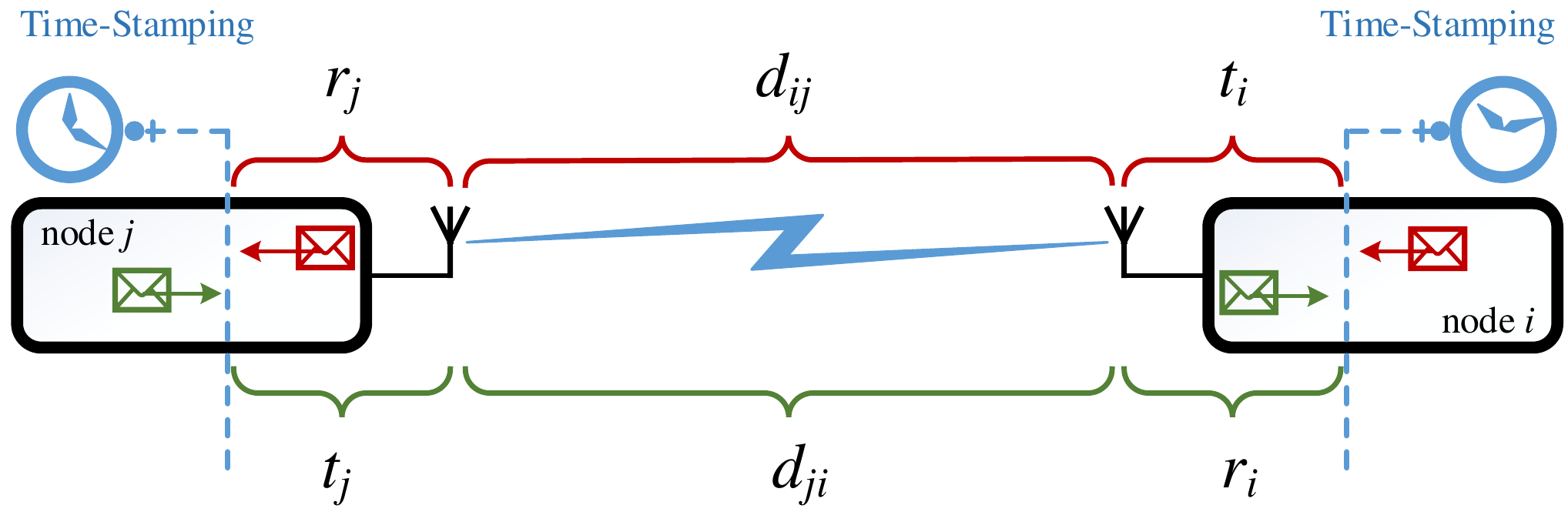}
\centering
\caption{Delay decomposition.}
\label{fig:deldec}
\end{figure}
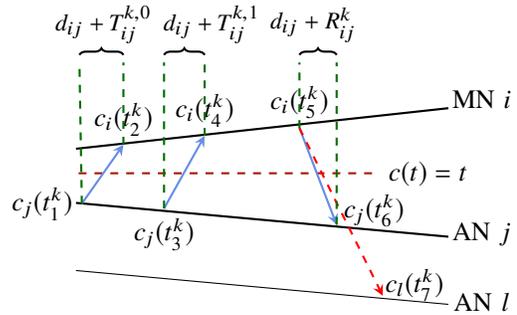
\begin{figure}[t!]
\begin{tikzpicture}[scale=.9]
	\definecolor{mc}{rgb}{0.4, 0.55, 0.9};
	\definecolor{mcshad}{rgb}{0.4, 0.55, 0.7};
	\definecolor{mc1}{rgb}{0.66, 0.11, 0.03};
	\definecolor{mc2}{rgb}{0.0, 0.44, 0.0};
	
	\draw (-3.5, 0.5) node(ci0){};
	\draw (-3.5, -0.3) node(cj0){};
	\draw (-3.5, -1.3) node(cj1){};
	\draw (2, 1.1) node(ci0end){};
	\draw (2.0, -0.8) node(cj0end){};
	\draw (2.0, -1.8) node(cj1end){};
	\draw[thick,-] (ci0.south)--(ci0end.south);
    \draw[thick,-] (cj0.south)--(cj0end.south);
    \draw[-] (cj1.south)--(cj1end.south);
    
	\draw (-3.6, 0) node(tbeg){};
	\draw (1.7, 0) node(tend){$c(t)=t$};
	\draw[mc1,thick,dashed] (tbeg.east)--(tend.west);
	
    \draw (-2.8, 0.8) node(cj2){$c_i(t_2^{k})$};
    \draw (-2.8, 0.8) node(cj2tex){};
    \draw (-0.2, 1.05) node(ci5) {$c_i(t_5^{k})$};
    \draw (-3.4, -0.8) node(ci1){};
    \draw (-4.0, -0.45) node(ci1){$c_j(t_1^{k})$ };
    \draw (1.1, -1.95) node(cl4){};
   	\draw (1.5, -1.65) node(cl4tex){$c_l(t_7^{k})$};
    \draw (0.35, -0.9) node(cj4){};
    \draw (0.9, -0.55) node(cj4tex){$c_j(t_6^{k})$};
    \draw (0, -0.85) node(cii){};
    \draw (-1.6, 0.93) node(ci4){$c_i(t_4^{k})$};
    \draw (-2.2, -0.9) node(cj3){$c_j(t_3^{k})$};
   
    \draw[mc,thick,-stealth] (cj3.north)--(ci4.south);
    \draw[mc,thick,-stealth] (ci1.east)--(cj2.south);
    \draw[thick, mc,-stealth] (ci5.south)--(cj4.north);
    
    \draw (-2.8, -1.35) node(cl2){};
    \draw (-1.6, -1.5) node(cl3){};
    \draw[red,thick, dashed,-stealth] (ci5.south)--(cl4);
    
    \draw (-3.45,1.5) node(b1){};
    \draw (-2.8,1.5) node(b2){};
    \draw (-1.6,1.5) node(b6){};
    \draw (-0.2,1.5) node(e1){};
    \draw (0.35,1.5) node(e2){};
    \draw (-2.2,1.5) node(e3){};
    \draw (-3.1,2.2) node(b3){\small $d_{ij}+T_{ij}^{k, 0}$};
    \draw (-0,2.2) node(b4){\small $d_{ij}+R_{ij}^k$};
    \draw (-1.5,2.2) node(b5){\small $d_{ij}+T_{ij}^{k, 1}$};
    \draw[thick,decorate,decoration={brace,raise=0.1cm}] (b1.north) -- (b2.north);
    \draw[mc2,thick,dashed] (cj2.south)--(b2.north);
    \draw[mc2,thick,dashed] (ci1.east)--(b1.north);
    \draw[thick,decorate,decoration={brace,raise=0.1cm}] (e1.north) -- (e2.north);
    \draw[mc2,thick,dashed] (cj4.north)--(e2.north);
    \draw[mc2,thick,dashed] (ci5.south)--(e1.north);
    
    \draw[mc2,thick,dashed] (cj3.north)--(e3.north);
    \draw[mc2,thick,dashed] (ci4.south)--(b6.north);
    \draw[thick,decorate,decoration={brace,raise=0.1cm}] (e3.north) -- (b6.north);
    
    \draw (2.5, 1.1) node(ci5) {MN $i$};
    \draw (2.5, -.9) node(ci5) {AN $j$};
    \draw (2.5, -1.9) node(ci5) {AN $l$};
\end{tikzpicture}
\centering
\caption{Time-stamp exchange between MN $i$ and AN $j$. Blue/red solid/dashed lines indicate the active/passive listening of AN $j$/$l$.}
\label{fig:stamp}
\end{figure}
\subsection{Offset Decomposition and Time-stamp Exchange}\label{ssec:offdec}
\subsubsection{Offset decomposition} To elaborate on the components making up the offset $\theta_i$, we break down this parameter as shown in Figure \ref{fig:deldec}. The parameter $t_j$/$t_i$ is the time taken for a packet to leave the transmitter after being time-stamped, $d_{ji}$/$d_{ij}$ denotes the propagation delay, and $r_i$/$r_j$ represents the time that a packet needs to reach the time-stamping point upon arrival at the receiver. In general, $t_j + d_{ji} + r_i \neq t_i + d_{ij} + r_j,$ indicating that the packets sent from node $j$ to node $i$ do not necessarily experience the same delay as those sent from node $i$ to node $j.$ Furthermore, we define $T_{ij} = t_j + r_i,$ and $R_{ij} = t_i + r_j$. Generally, $T_{ij}$ and $R_{ij}$ (and correspondingly $t_j,$ $t_i,$ $r_j,$ and $r_i,$) are random variables due to several hardware-related random independent processes and can, therefore, be assumed i.i.d. Gaussian random variables, whereas $d_{ji}$ and $d_{ij}$ are usually assumed to be deterministic and symmetric ($d_{ji} = d_{ij}$) \cite{wu2010clock,leng2011distributed}.
The random variables $T_{ij}$ and $R_{ij}$ are assumed to be distributed as $\mathcal{N}(\mu_T, \sigma^2_T)$ and $\mathcal{N}(\mu_R, \sigma^2_R),$ respectively. 
As mentioned in \cite{leng2011distributed,du2013distributed,etzlinger2014cooperative}, while it is typical to assume that $\mu_T = \mu_R,$ and parameters $\sigma_{T}$ and $\sigma_{R}$ are known,
having any information about the value of $\mu_T$ and $\mu_R$ is highly unlikely. Therefore, we construct the joint sync\&loc algorithm assuming no knowledge on $\mu_T$ and $\mu_R$ except for $\mu_T = \mu_R.$
\subsubsection{Time-stamp exchange scheduling}\label{sssec:schedule}
We deploy the asymmetric time-stamp exchange mechanism shown in Figure \ref{fig:stamp}, proposed in \cite{chepuri2012joint} and employed in \cite{etzlinger2014cooperative,etzlinger2017cooperative}.
The AN $j$ propagates a message announcing the beginning of a time-stamp exchange round. Upon reception, the connected MNs go to \textit{active listening} mode while the neighboring ANs go into \textit{passive listening} mode. In the former, the MNs will respond after reception of two messages from AN $j$ (depicted in Figure \ref{fig:stamp}), whereas, in the latter, the ANs only listen to the packet exchange between AN $j$ and MNs. Without loss of generality and for the sake of simplicity we write the equations for only one MN and two ANs. The extension to multiple ANs/MNs is straightforward. 

\subsubsection{Time-stamp exchange mechanism}
Given Section \ref{sssec:schedule}, and considering AN $j$ as master node\footnote{In Figure \ref{fig:stamp}, instead of a global time reference $c(t)=t,$ we take node $j$ as master node. It is straightforward to see that $\frac{1}{\tilde{\gamma}_i}=\frac{\gamma_j}{\gamma_i},$ $\tilde{\theta}_i = \theta_i-\tg\theta_j,$ $\tilde{d}_{ij} + \tijt{k} = \gamma_j(d_{ij} + \tij{k}),$ and $\tilde{d}_{ij} - \rijt{k} = \gamma_j(d_{ij} - \rij{k})$. For the sake of simplicity, as done in \cite{wu2010clock}, we assume $\tilde{d}_{ij}=d_{ij},$ $\rijt{k} = \rij{k},$ and $\tijt{k}=\tij{k}.$ This is valid because $\gamma_j\approx 1$ and the values of $d_{ij} + \tij{k}$ and $d_{ij} - \rij{k}$ are low.}, we can write
\begin{align}
&\tgam(c_i(t_{2}^k) - \tilde{\theta}_i) = c_j(t_1^k) + \frac{d_{ij}}{v_c} + \tij{k, 0}  ,\label{eq:c1}\\ 
&\tgam(c_i(t_{4}^k) - \tilde{\theta}_i) = c_j(t_3^k) + \frac{d_{ij}}{v_c} + \tij{k, 1}, \label{eq:c2}\\
&\tgam(c_i(t_{5}^k) - \tilde{\theta}_i) = c_j(t_6^k) - \frac{d_{ij}}{v_c} - \rij{k}, \label{eq:c3} 
\end{align}
where $t_1^k$/$t_2^k$, $t_3^k$/$t_4^k$, and $t_5^k$/$t_6^k$ are the time points where MN $i$ and AN $j$ send/receive the sync messages, respectively. Parameter $d_{ij} = \dij{j}{}$ denotes the Euclidean distance between nodes $i$ and $j$ and $v_c$ is the speed of light.
Furthermore, if there is an AN $l$ in passive listening mode, we can write
\begin{align}
&\tgam(c_i(t_{5}^k) - \tilde{\theta}_i) = c_l(t_7^k) + \theta_{jl} - \frac{d_{il}}{v_c} - R_{il}^k, \label{eq:c4}
\end{align}
where $t_7^k$ is the time point when AN $l$ receives the time-stamps sent by MN $i$. Parameter $\theta_{jl}$ denotes the relative offset between ANs $j$ and $l$ and is shown in \cite{goodarzi2020synchronization} to have the distribution $\mathcal{N}(\theta_{jl}|0,\sigma_{jl}^2)$ with $\sigma_{jl}\approx 1$ ns for an urban scenario similar to Figure \ref{fig:scenario}. Note that we neglect the impact of skew difference between $l$ and $j$ since it has been shown that this difference is almost zero if the ANs frequently synchronize to the backhauling BS using the algorithm introduced in \cite{goodarzi2020hybrid}.

At the $k$-th round of time-stamp exchange (and correspondingly $k$-th round of joint sync\&loc), the network localization center is expected to have collected the time-stamps  
$$\ccij^k = \left[c_j(t_1^k), c_i(t_2^k), c_j(t_3^k), c_i(t_4^k), c_i(t_5^k), c_j(t_6^k), c_l(t_7^k)\right]. $$

\subsection{Angle of Arrival}\label{sssec:crb}
AoA estimation has been extensively investigated in the literature. 
In particular, beamforming, subspace, and maximum likelihood methods can be employed to accurately estimate the AoA \cite{grosssmart}. Nevertheless, in this work, our focus is to reveal the potential merit of time-stamp exchange in joint sync\&loc. Therefore, we assume that an uncertain estimation of AoA is available where the uncertainty, i.e. $\sigma_{\varphi},$ is obtained from the CRB. 

Assuming that each AN has a $N$-element Uniform Linear Array (ULA) antenna,
the CRB on AoA estimation can then be given by \cite{fittipaldi2008cramer}
\begin{equation}
J(\varphi)^{-1} = \left(\frac{N(N-1)(N+1)\pi^2\sin^2(\varphi)}{24}\times \text{SNR}\right)^{-1}.
\end{equation}
We set the maximum value of SNR to 30 dB which occurs at the closest MN-AN distance of 5m. It then drops according to Friis path loss formula, i.e. $20\log_{10}(d_{ij})$. Furthermore, the number of AN antennas, $N,$ and distance between them are set to $16$ and $\frac{\lambda}{2}$, respectively, where $\lambda$ denotes the wavelength.
Thus, at the $k$-th round of time-stamp exchange, each AN is expected to have estimated $\varphi^k$, which, in this work, is derived from the distribution $\mathcal{N}(\varphi_{p}^k, \sigma_{\varphi}^2)$ with $\sigma_{\varphi}^2 = J(\varphi)^{-1}$ and $\varphi_{p}^k$ being calculated knowing the exact location of the MN $i$ and AN $j$.
\section{Clock Parameters and Position Estimation}
\begin{figure}[t!]
\begin{tikzpicture}[scale=1]
    \draw (0, 0) node[inner sep=0] {\includegraphics[width=0.85\linewidth]{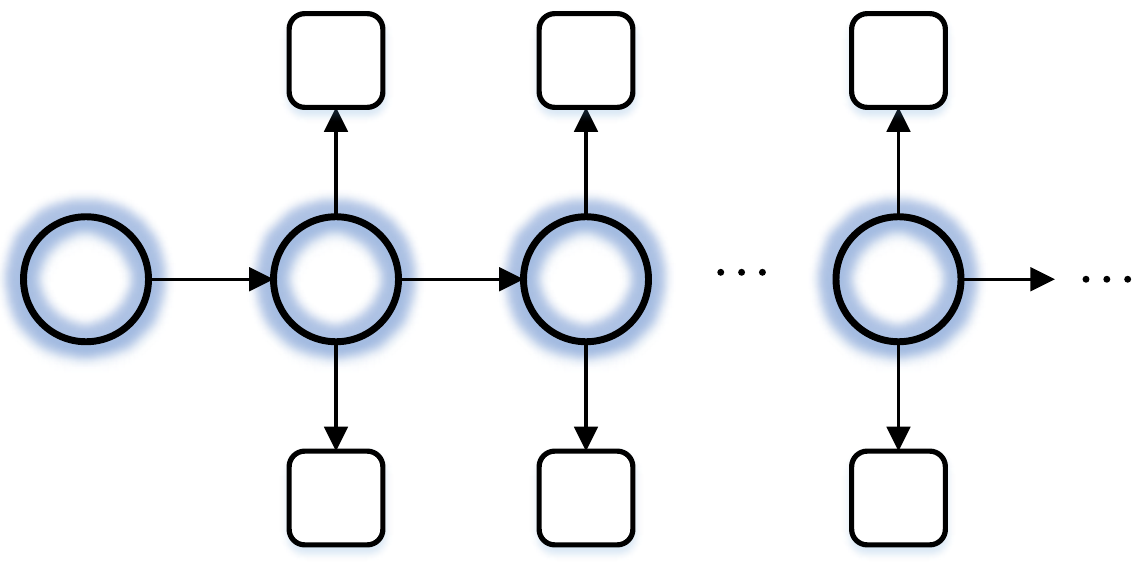}};
	\draw (-3.2,0.0) node(a00){$\tetvect{i}^0$};  
	\draw (-1.55,0.0) node(a00){$\tetvect{i}^1$};
	\draw (.1,0.0) node(a00){$\tetvect{i}^2$};
	\draw (2.15,0.0) node(a00){$\tetvect{i}^k$};
	
	\draw (-1.55,-1.45) node(a00){$\ccij^1$};
	\draw (.1,-1.45) node(a00){$\ccij^2$};
	\draw (2.15,-1.45) node(a00){$\ccij^k$};
	
	\draw (-1.55,1.45) node(a00){$\varphi^1$};
	\draw (.1,1.45) node(a00){$\varphi^2$};
	\draw (2.15,1.45) node(a00){$\varphi^k$};
\end{tikzpicture}
\centering
\caption{Representation of Bayesian estimation.}
\label{fig:bayesrep}
\end{figure}
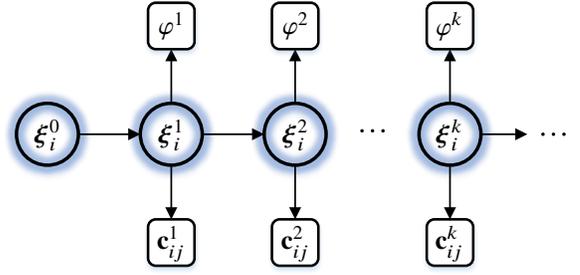

Let $\ttat{k}$ be the state of the vector variable $\tetvect{i} \triangleq \begin{bmatrix}\tgam & \frac{\ttet}{\tg} & x_i & y_i & v_{x_i} & v_{y_i}\end{bmatrix}^T$ after the $k$-th round of time-stamp exchange (visualized in Figure \ref{fig:bayesrep}). Parameters $x_i$/$v_{x_i}$ and $y_i$/$v_{y_i}$ denote the position/velocity of node $i$ on the $x$ and $y$ axis, respectively.  
The \textit{probability distribution function} (pdf) corresponding to  the $k$-th state can then be written as
\begin{equation}
 p(\ttat{k}|\ccij^{1:k}, \varphi^{1:k}) = \int p(\ttat{0},\cdots, \ttat{k}|\ccij^{1:k}, \varphi^{1:k})\ d\Theta^{k-1},
 \label{eq:pdfk}
\end{equation}
where $\Theta^{k-1} = \left[\ttat{0},\cdots,\ttat{k-1}\right]$. Applying Bayesian rule, we can rewrite (\ref{eq:pdfk}) as
\begin{multline}
p(\ttat{k}|\ccij^{1:k}, \varphi^{1:k}) \propto \\ \int p(\ccij^{1:k}, \varphi^{1:k}|\ttat{0},\cdots, \ttat{k})p(\ttat{0},\cdots, \ttat{k})\ d\Theta^{k-1}.
\label{eq:bayesrule}
\end{multline}
Knowing that the measurements are independent and assuming Markov property \cite{barker1995bayesian}, the integrands in (\ref{eq:bayesrule}) can be reformulated as
\begin{align}
&p(\ccij^{1:k}, \varphi^{1:k}|\ttat{0},\cdots, \ttat{k}) = \cntnt{k}{k}\cdots \cntnt{1}{1},& \nonumber\\
&p(\ttat{0},\cdots, \ttat{k}) = \tntt{k}{k-1}\cdots \tntt{1}{0}p(\tetvect{i}^0),&
\label{eq:markov}
\end{align}
where $p(\tetvect{i}^0)$ denotes the prior knowledge on $\tetvect{i}.$ Plugging (\ref{eq:markov}) into (\ref{eq:bayesrule}) and carrying out mathematical simplifications as in \cite{barker1995bayesian,goodarzi2020synchronization,goodarzi2020hybrid} leads to
\begin{align}
p(\ttat{k}|\ccij^{1:k}, \varphi^{1:k}) \propto p(\tetvect{i}^{k}|\ccij^{1:k-1}, \varphi^{1:k-1})\cntnt{k}{k}.
\label{eq:bayesfin}
\end{align}
The term $p(\tetvect{i}^{k}|\ccij^{1:k-1}, \varphi^{1:k-1})$ is referred to as \textit{prediction} step while the term $\cntnt{k}{k}$ is considered as \textit{correction} step \cite{barker1995bayesian}. In wireless networks, it is typical to assume that $\tetvect{i}^k$ is Gaussian distributed \cite{wu2010clock,etzlinger2014cooperative,etzlinger2017cooperative}. Given this assumption, if the relation between all the states in Figure \ref{fig:bayesrep} is linear, we can conclude that the marginal in (\ref{eq:bayesfin}) is also Gaussian distributed. While that is the case for the prediction step, the measurement equations (and consequently, the correction steps) are non-linear, and therefore, need to be linearized. In the sequel, we deal with the details of prediction and correction steps.
\subsubsection{Prediction} Given the dynamics of MNs' clocks and movements, a reasonable prediction for $\tetvect{i}^k$ is given by \cite{etzlinger2017cooperative},  

\begin{equation}
\tetvect{i}^k = \mathbf{A}\tetvect{i}^{k-1}+\mathbf{n}^{k-1}_i,
\label{eq:predlin}
\end{equation}
where $$\mathbf{A}=\begin{bmatrix} \mathbf{I}_2 & \mathbf{0}_2 & \mathbf{0}_2 \\
   \mathbf{0}_2 & \mathbf{I}_2 & \Delta\mathbf{I}_2 \\
   \mathbf{0}_2 & \mathbf{0}_2 & \mathbf{I}_2
  \end{bmatrix}.$$
Parameter $\Delta$  is the time difference between two consecutive rounds of time-stamp exchange and $\mathbf{n}^{k-1}_i$ denotes the Gaussian noise vector and assumed to have zero mean and covariance matrix\footnote{In general, design of $\mathbf{Q}_n$ is a difficult task. In particular, if it is too small, the filter will be overconfident in its prediction model and will diverge from the actual solution. In contrast, if it is too large, then it will be unduly dominated by the noise in the measurements and perform sub-optimally. In this work, we follow the design model discussed in \cite{labbe2019kalman,khan2014localization}.} $\mathbf{Q}_n = \diag(\sigma_{\gamma}^2, \sigma_{\theta}^2, \sigma^2_x,\sigma^2_y, \sigma_{v_x}^2, \sigma_{v_y}^2)$. 
Given (\ref{eq:predlin}), the prediction term can be written as
\begin{equation}
 p(\tetvect{i}^{k}|\ccij^{1:k-1}, \varphi_{ij}^{1:k-1}) \sim \mathcal{N}(\tetvect{i}^k | \mupred, \copred),
 \label{eq:predpdf}
\end{equation}
where $\mupred  = \mathbf{A}\boldsymbol{\mu}_i^{k-1} $ and $\copred = \mathbf{A}\mathbf{\Sigma}^{k-1}_i\mathbf{A}^T + \mathbf{Q}_n.$
\subsubsection{Correction} We conduct the following mathematical manipulations to obtain the correction term in (\ref{eq:bayesfin}). Subtracting (\ref{eq:c1}) from (\ref{eq:c2}) leads to
\begin{align}
& \frac{1}{\tilde{\gamma}_i}(c_i(t_{4}^{k}) - c_i(t_{2}^{k})) = c_j(t_3^{k}) - c_j(t_1^{k, 0}) + \tij{k, 1}-\tij{k, 0}, \label{eq:c2-c1}
\end{align} 
while summing up (\ref{eq:c2}) and (\ref{eq:c3})
\begin{align}
&\tgam(c_i(t_{4}^k) + c_i(t_{5}^k)-2\ttet) = c_j(t_3^k) + c_j(t_6 ^k) + \tij{k, 1}-\rij{k}. \label{eq:c2+c3} 
\end{align}
Equation (\ref{eq:c3}) stays as it is unless there are extra ANs cooperating with AN $j$ by passively listening to the time-stamp exchange. For example, for one extra AN cooperating with AN $j,$
subtracting (\ref{eq:c3}) from (\ref{eq:c4}) provides
\begin{align}
&\frac{d_{il}-d_{ij}}{v_c} =  c_l(t_7^k) - c_j(t_6^k) - \theta_{jl} + \rij{k} - R_{il}^k.
\label{eq:c4-c3} 
\end{align}
Finally, the AoA measurement can be expressed as follows:
\begin{align}
\arctan(\frac{y_i-y_j}{x_i-x_j}) = \varphi_j^k \label{eq:aoa} 
\end{align} 
where $\varphi_i^k$ is calculated as explained in Section \ref{sssec:crb}. Again, if there are more ANs involved in joint sync\&loc, one can write the same equation for their AoA measurements.

 
To permit (\ref{eq:bayesfin}) to have a closed-form solution, the relation between parameters in the measurement equations (\ref{eq:c2-c1}), (\ref{eq:c2+c3}), (\ref{eq:c3}), and (\ref{eq:aoa}) must be linear. However, this is not the case as the distance function is not linear. Therefore, we draw on Taylor expansion to linearize the non-linear terms, thereby allowing for a closed-form solution for (\ref{eq:bayesfin}). In particular, \textit{we write the Taylor expansion around the point predicted by the prediction step} in (\ref{eq:predlin}). Thus
%
\begin{align}
&\frac{d_{ij}}{v_c} \approx a^k_{j,0} + a_{j,x}^k (x_i-x_i^{k}) + a_{j,y}^k (y_i-y_i^{k}),
\label{eq:taylor1} \\
&\arctan(\frac{y_i-y_j}{x_i-x_j}) \approx b_{j,0}^k + b^k_{j,x} (x_i-x_i^{k}) + b^k_{j,y} (y_i-y_i^{k}), \label{eq:taylor2}
\end{align}
\begin{figure*}
\begin{align}
&a^k_{j,0} = \frac{1}{v_c}\left(\dij{j}{k}\right),  && a_{j,x}^{k} = \frac{x_i^{k}-x_j}{v_c^2 a^k_{j,0}},   && a_{j,y}^{k} = \frac{y_i^{k}-y_j}{v_c^2 a^k_{j,0}}, \label{eq:constay1}\\ 
&b^k_{j,0} = \arctan(\frac{y^{k}_i-y_j}{x^{k}_i-x_j}), && b_{j,x}^k =- \frac{y_i^{k}-y_j}{v_c^2(a^k_{j,0})^2}, && b_{j,y}^k = \frac{x_i^{k}-x_j}{v_c^2(a^k_{j,0})^2}.\label{eq:constay2}
\end{align}
\hrule
\end{figure*}
with $a^k_{j,0},$ $a^k_{j,x},$ $a^k_{j,y},$ $b^k_{j,0},$ $b^k_{j,x},$ and $b^k_{j,y},$ calculated as in (\ref{eq:constay1}) and (\ref{eq:constay2}). 
Given (\ref{eq:taylor1}) and (\ref{eq:taylor2}), and computing the average velocity using 
\begin{align}
v_{x_i} = \frac{x_i - x_i^{k-1}}{\Delta}, && v_{y_i} = \frac{y_i - y_i^{k-1}}{\Delta},
\end{align}
we can write (\ref{eq:c2-c1}), (\ref{eq:c2+c3}), (\ref{eq:c3}), and (\ref{eq:aoa}) for single-AN localization in matrix form as
\begin{align}
\mathbf{B}_{ij}\tetvec{i} = \mathbf{r}_{ij} + \zji,
\end{align}
where $\zji\sim \mathcal{N}(\mathbf{z}|\mathbf{0},\mathbf{R}_{ij})$ with $$\mathbf{R}_{ij} = \diag(2\sigma^2_{T_{ij}}, \sigma^2_{T_{ij}} + \sigma^2_{R_{ij}}, \sigma^2_{R_{ij}}, \sigma_{\varphi}^2, (\frac{\sigma^{k-1}_{x_i}}{\Delta})^2, (\frac{\sigma^{k-1}_{y_i}}{\Delta})^2),$$
 $$ \mathbf{B}_{ij} = \begin{bmatrix} 
 \begin{matrix}
 c_i(t_{4}^{k}) - c_i(t_{2}^{k}) & 0 \\
 c_i(t_{4}^k) + c_i(t_{5}^k) & -2 
 \end{matrix}
 & \mathbf{0}_2 & \mathbf{0}_2\\ \begin{matrix}
 c_i(t_{5}^k) & -1 \\ 0 & 0
\end{matrix}  & 
 \begin{matrix}
 a^k_{j,x} & a^k_{j,y} \\ b_{j,x}^k & b^k_{j,y} \\
 \end{matrix} & \mathbf{0}_2 \\
 \mathbf{0}_2 & -\frac{1}{\Delta}\mathbf{I}_2 & \mathbf{I}_2
\end{bmatrix},$$
and $\mathbf{r}_{ij}$ is constructed as in (\ref{eq:rij}).
\begin{figure*}[t]
\begin{align}
\mathbf{r}_{ij} = \left[
c_j(t_3^{k}) - c_j(t_1^{k}), c_j(t_{3}^k) + c_j(t_{6}^k), c_j(t_{6}^k) - a_{j,0}^k + a_{j,x}^k x_i^{k} + a_{j,y}^k y_i^{k}, \varphi_j^k - b_{j,0}^k + b_{j,x}^k x_i^{k} + b_{j,y}^k y_i^{k}, -\frac{x_i^{k-1}}{\Delta}, -\frac{y_i^{k-1}}{\Delta}
\right]^T.
\label{eq:rij}
\end{align}
\hrule
\end{figure*}
The extension to two-AN localization can be readily carried out by a) replacing (\ref{eq:c3}) with (\ref{eq:c4-c3}), b) writing an extra equation similar to (\ref{eq:aoa}) for AN $l$, c) changing the $\mathbf{B}_{ij},$ $\mathbf{r}_{ij},$ and $\mathbf{R}_{ij}$ accordingly.
 Finally, the correction term can be written as
\begin{equation}
\cntn{k}{k} \sim \mathcal{N}(\mucorr, \cocorr),
\label{eq:mespdf}
\end{equation}
where $\mucorr = (\mathbf{B}_{ij}^T\mathbf{B}_{ij})^{-1} \mathbf{B}_{ij}^T \mathbf{r}_{ij},$ and $$\cocorr = (\mathbf{B}_{ij}^T\mathbf{B}_{ij})^{-1}\mathbf{B}_{ij}^{T}\mathbf{R}_{ij}\mathbf{B}_{ij}(\mathbf{B}_{ij}^T\mathbf{B}_{ij})^{-T}.$$
\subsubsection{Estimation} Considering (\ref{eq:predpdf}) and (\ref{eq:mespdf}), the estimated distribution in (\ref{eq:bayesfin}) is given by
\begin{equation}
p(\ttat{k}|\ccij^{1:k}, \varphi_{ij}^{1:k})\sim \mathcal{N}(\muest, \coest),
\label{eq:estpdf}
\end{equation}
where
\begin{align}
&\muest = \left[\copred + \cocorr\right]^{-1}\left(\cocorr\mupred + \copred\mucorr\right), \\
&\coest = \left[\copred^{-1} + \cocorr^{-1}\right]^{-1}.
\end{align}
The parameters in (\ref{eq:predpdf}), (\ref{eq:mespdf}), and (\ref{eq:estpdf}) are calculated recursively and, in each iteration $k,$ the estimation of the clock skew, clock offset, and position can be obtained by
\begin{equation}
\tg^k = \frac{1}{\muest(1)},\ \ttet^k = \frac{\muest(2)}{\muest(1)},\ x_i^k = \muest(3),\ \text{and}\ y_i^k = \muest(4).
\label{eq:finestpair}
\end{equation}
Algorithm \ref{alg:brf} summarizes this recursive process.

It is worth mentioning that the position initialization has a major impact on the performance of the algorithm and can, if inappropriately chosen, lead to its divergence. In this work, similar to \cite{werner2015joint}, we assume that the initial position of the MN is available via Global Navigation Satellite System (GNSS). The initialization of clock parameters is, however, straightforward and can be done, according to \cite{goodarzi2020synchronization,goodarzi2020hybrid,du2013distributed}, with $\mathcal{N}(1, \infty)$ and $\mathcal{N}(0, \infty)$ for clock skew and offset, respectively.
\begin{algorithm}[t!]
\begin{algorithmic}[1]
\STATE Initialize p($\tetvect{i}^0$) using information about MN position available via, e.g., GNSS\label{init}
\WHILE {MN is in LoS of AN $j$}  \label{forBRF}
\STATE Calculate the mean vector and covariance matrix of the \textit{prediction} pdf using (\ref{eq:predpdf}) \label{predpdf}
\STATE Perform the time-stamp exchange mechanism described in Section \ref{sssec:schedule} and Figure \ref{fig:stamp}
\STATE Construct $\mathbf{B}_{ij},$ $\mathbf{R}_{ij},$ and $\mathbf{r}_{ij}$ using the measurements and obtain the mean vector and covariance matrix of \textit{correction} pdf using (\ref{eq:mespdf})\label{mespdf}
\STATE Compute the mean vector and covariance matrix of the estimation $\tetvect{i}^k$ using (\ref{eq:estpdf})\label{finpdf}
\ENDWHILE \label{endfor}
\end{algorithmic}
\caption{BRF-based joint sync\&loc}
\label{alg:brf}
\end{algorithm}

%
\section{Simulation Results and Discussion}
We perform analysis for two scenarios shown in Figure \ref{fig:scenario}, which are regarded in \cite{werner2015joint} as challenging.
In scenario (a), a pedestrian moves with a constant velocity of $2$ m/s ($\approx$7 km/h) and takes the turns randomly until it exits the map.
In scenario (b), a car commences its journey by accelerating to reach the velocity of $14$ m/s ($\approx 50$ km/h). It continues moving with constant velocity and decelerates upon approaching the intersection until it completely stops (e.g. due to the red light). The same repeats between two intersections. At the the second intersection, it begins moving and takes the turn and continues to accelerate to $14$ m/s limit until it exits the map. All the turns as well as acceleration (acc.) coefficients are chosen randomly. Moreover, the Root Mean Square Error (RMSE) obtained by \cite{werner2015joint} (i.e. 3m/0.5m and 10ns/4ns for position and clock offset estimation of 1-AN/2-AN, respectively) serves as the baseline to our approach. Nevertheless, \cite{werner2015joint} does not address the impact of $\mu_T$ and variable velocity (scenario (b)). Finally, additional simulation parameters can be found in Table \ref{tab:sim}.
\begin{table}[t!]
\centering
\caption{Simulation parameters}
\begin{tabular}{lc}
Parameters & Values \\ \hline
\# of independent simulations &	1000 \\ 
Initial random delays ($\tilde{\theta}_i$) &	[-1000, 1000] ns \\ 
Random acc. range & $\pm$[1, 2.5] m/s$^2$ \\
STD of acc. noise ($\sigma_{a_x}, \sigma_{a_y}$) & 2.5 m/s$^2$ \\ 
Period of joint sync\&loc ($\Delta$) & 200 ms \\
Process noise covariance matrix ($\mathbf{Q}_n$) & $\diag(10^{-12}, 10^{-2}, (0.5\sigma_{a_x}\Delta)^2,$ \\  
& $(0.5\sigma_{a_y}\Delta)^2, \sigma_{a_x}^2, \sigma_{a_y}^2)$ \\ 
Max. velocity for scenarios (a), (b) & 2, 14 (m/s) \\
AN density & 50 m \\ \hline
\end{tabular}
\label{tab:sim}
\end{table}

\begin{figure}[t!]
\includegraphics[width=0.95\linewidth]{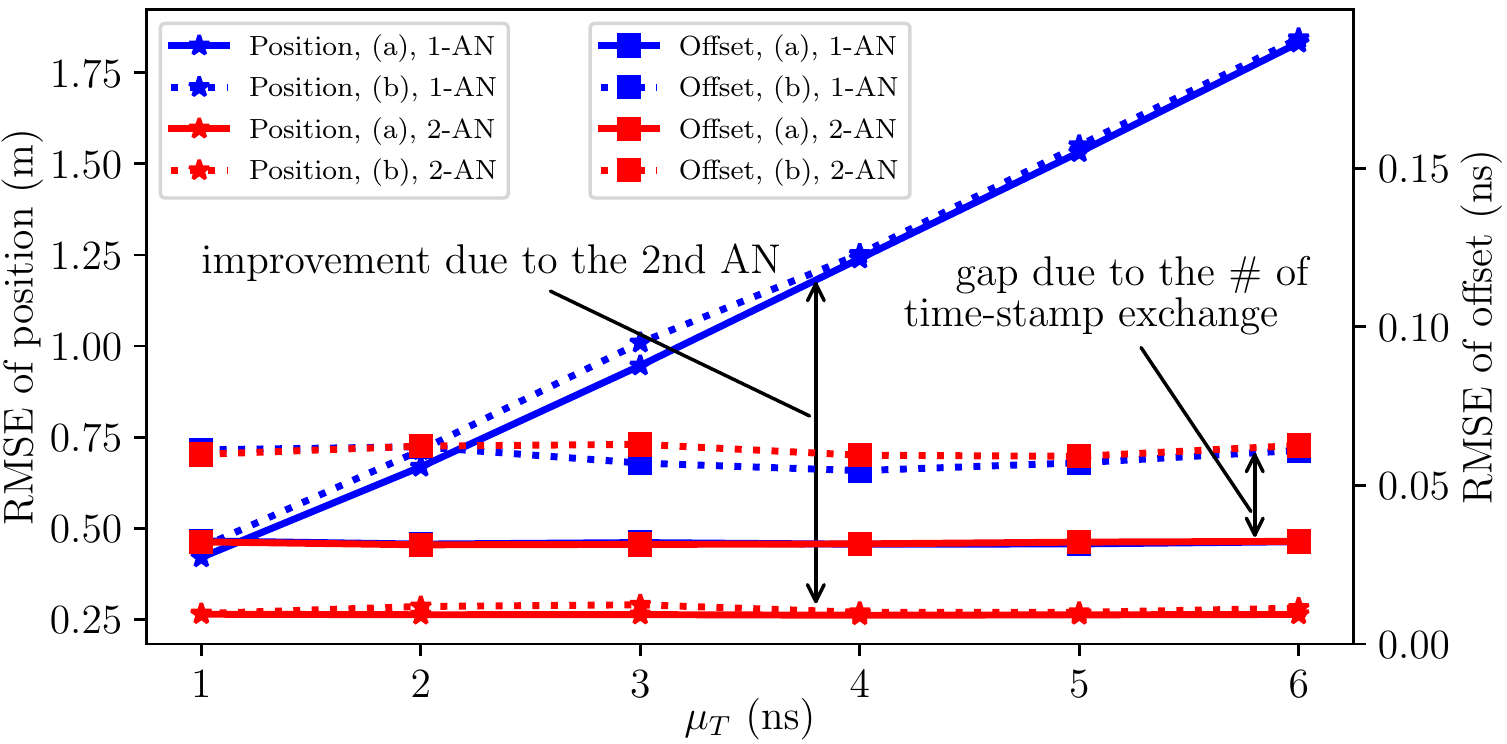}
\centering
\caption{Performance of joint sync\&loc algorithm ($\sigma_{T}=0.2$ns). Slope of increase in RMSE of position for 1-AN case = $0.28$ m/ns.}
\label{fig:perform1}
\end{figure}

Figure \ref{fig:perform1} shows the RMSE of position and clock offset estimation with respect to $\mu_{T}$ (or alternatively $\mu_{R}$) for $\sigma_T = \sigma_R = 0.2$ ns. As can be seen, the RMSE of position increases for the single AN (1-AN) as $\mu_T$ grows whereas it remains almost unchanged for the two AN (2-AN) case. The reason is disguised in (\ref{eq:c3}) and (\ref{eq:c4-c3}). In the former, the position parameters are impaired by random variable $R_{ij} \sim \mathcal{N}(\mu_T, \sigma_T^2)$ while in the latter by $(R_{ij}-R_{il})\sim \mathcal{N}(0, 2\sigma_T),$ which is obviously a zero mean Gaussian variable and, therefore, indifferent to the growth of $\mu_T.$ It is clear that if $\mu_R$ is not equal for the two ANs (e.g. they feature different hardware), the RMSE of the 2-AN case would increase as well, albeit with a smaller slope than \mbox{1-AN} case.  Furthermore, for the same reason, the RMSE of the clock offset estimation remains almost constant with the increase of $\mu_T$. Moreover, the gap between RMSE of the clock offset estimation in two scenarios is due to the higher number of time-stamp exchanges in (a) where the journey takes longer, given the constant velocity of 2 m/s.  

\begin{figure}[t!]
\includegraphics[width=0.9\linewidth]{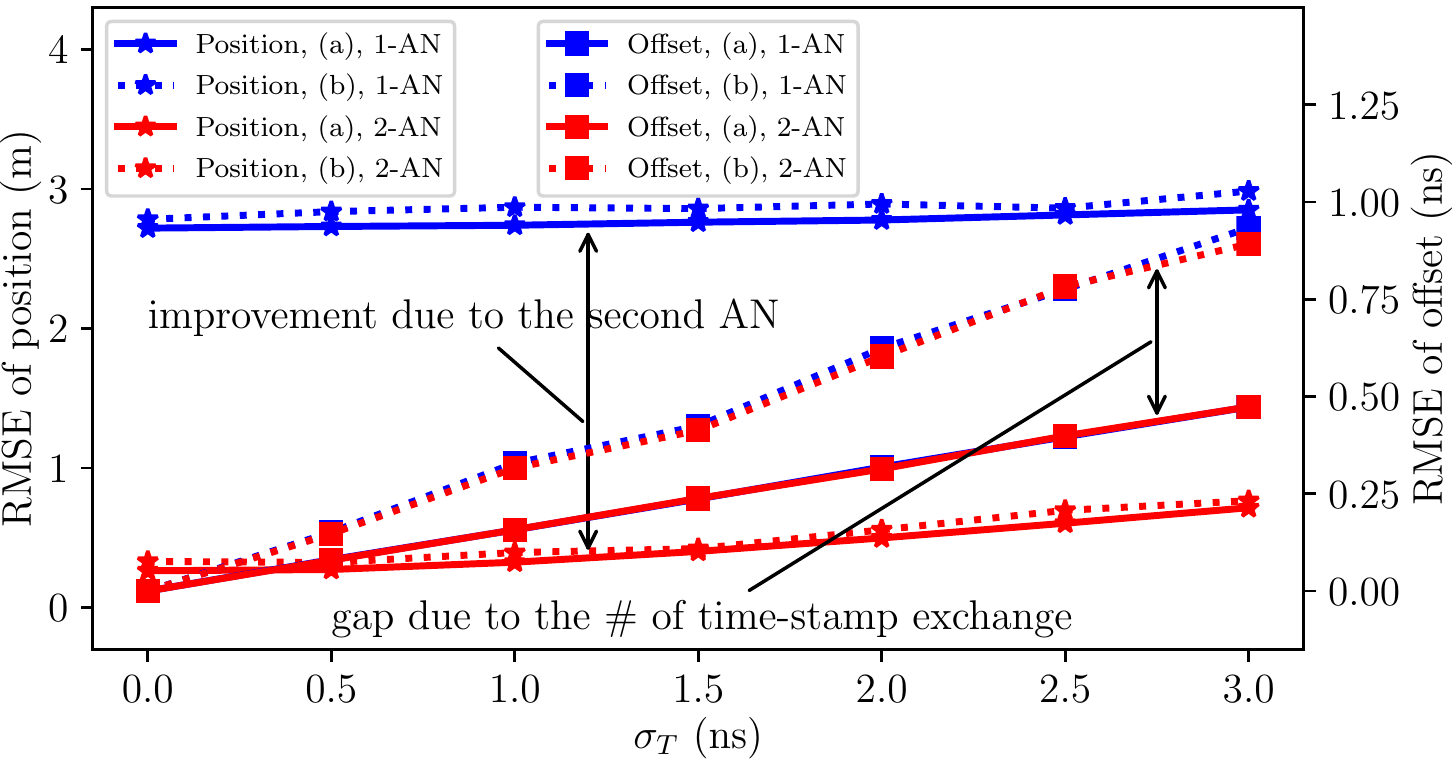}
\centering
\caption{Performance of joint sync\&loc algorithm ($\mu_{T}=9$ns). Slope of increase in RMSE of position for 2-AN case = $0.15$ m/ns.}
\label{fig:perform2}
\end{figure}

Figure \ref{fig:perform2} presents the RMSE of position and clock offset estimation versus $\sigma_{T}$ for $\mu_T = \mu_R = 9$ ns. It can be noticed that the RMSEs of position and clock offset grow with the increase of uncertainty in time-stamps. In particular, the growth rate in RMSE of position is higher for 2-AN case as the uncertainty in (\ref{eq:c3}) differs from that of (\ref{eq:c4-c3}) by factor of two. In fact, this growth for 1-AN case is very smooth that we can consider it as negligible. Moreover, the RMSE of the clock offset estimation increases for both 1-AN and 2-AN in both scenarios as (\ref{eq:c2-c1}) and (\ref{eq:c2+c3}) are identical in all the cases. Again, the gap between RMSE of clock offset estimation in two scenarios is due to higher number of time-stamp exchanges in (a). 

Considering both Figures, we can remark that while uncertainty in time-stamping, i.e. $\sigma_T$ and $\sigma_R,$ can be alleviated relatively well using BRF (especially for 1-AN case), the delay in time-stamping, i.e. $\mu_T$ and $\mu_R,$ can only be mitigated by either employing multiple ANs or improving the hardware responsible for time-stamping. In particular, for sub-meter accuracy localization via a single AN, the time-stamping mechanism should be designed such that $\mu_T$ is kept below 3ns.
\section{Conclusion and Future work}
We presented an algorithm for joint sync\&loc of mobile users in communication networks. In particular, we leveraged on asymmetric time-stamp exchange, which is traditionally utilized for time synchronization, to estimate clock offset and skew while simultaneously obtaining information about the distance between ANs and MNs. Further on, we combined the aforementioned information with AoA estimation to localize the MNs. Simulation results indicate that while the performance of the proposed algorithm is promising, the position and clock offset estimation errors are highly dependent on the delay in hardware time-stamping as well as its accuracy. We mitigated the negative impact of this dependency by deploying more ANs for performing joint sync\&loc. 

In this work, we drew on CRB of AoA to carry out simulations. However, in practice, AoA estimation can be challenging and impose limitation on the performance of the algorithm. Therefore, in the future works, we will employ a suitable AoA estimation algorithm and the hardware at our disposal to evaluate the performance of our proposed joint sync\&loc algorithm in practice. 

\bibliography{synch_loc_paper_2}
\bibliographystyle{IEEEtran}
\end{document}